\documentclass{kapproc}
\usepackage{procps} 
\usepackage[dvips]{graphicx}
\usepackage{psfig}
\upperandlowercase
 \let\footnote\savefootnote
 \let\footnotetext\savefootnotetext

\normallatexbib 

\begin{document}

\articletitle
{Radio Polarimetry in the Southern \\ Galactic Plane Survey}

\chaptitlerunninghead{Southern Galactic Plane Survey} % Shorter running head title.

 \author{M. Haverkorn\altaffilmark{1}, B.~M. Gaensler\altaffilmark{1},
   N.~M. McClure-Griffiths\altaffilmark{2},
   J.~M. Dickey\altaffilmark{3},  A.~J. Green\altaffilmark{4}}

\affil{\altaffilmark{1}Harvard-Smithsonian Center for Astrophysics
       \altaffilmark{2}Australia Telescope National Facility, CSIRO
       \altaffilmark{3}Dpt.\ of Astronomy, Univ.\ of Minnesota
       \altaffilmark{4}Astrophysics Dpt., Univ.\ of Sydney}

\mbox{}\\
We present radio polarimetric observations of a region spanning
$325.5^\circ<l<332.5^\circ$, $-0.5^\circ<b<3.5^\circ$, used as a test region
for the Southern Galactic Plane Survey (SGPS) \cite{GDM01}.
The SGPS is a radio survey in the Southern Galactic plane in H{\sc I}
and polarized continuum at 9~frequencies around 1.4~GHz, performed with the
Australia Telescope Compact Array (ATCA) and the Parkes 64m dish\cite{MGD01}. 
Ubiquitous structure in linearly polarized intensity is often
uncorrelated with structure in total intensity, indicating Faraday
rotation and depolarization. 

The second order structure function (SF) of RM $SF_{RM}({\mathbf r}) =
\langle (RM({\bf x}) - RM({\bf x+r}))^2 \rangle_{\bf x}$ (where
$\langle\rangle_{\bf x}$ denotes averaging over all positions ${\bf
  x}$ in the field) for different position angles $\alpha$ of the
vector ${\bf r}$  shows a very shallow slope of the SF ($\sim$0.2) on
small scales and an anisotropy in the slope on large scales.The
anisotropy on large scale scales is indicative of a gradient or large
filament across the field. 

The shallow slope can be explained by contributions of two separated
Faraday screens. Then, the saturation scale of the SF denotes the
outer scale of structure in the nearest Faraday screen. Assuming that
the two Faraday screens are the local and Carina spiral arms, at
distances of $\sim$100~pc and $\sim$1500~pc, respectively, the
saturation point of the SF at log(r) = $-0.2$ denotes the outer scale
of structure in the local arm, i.e. about 2~pc. This scale is similar
to the typical scale of a Str\"omgren sphere of a B1-2
star. Therefore, we suggest that H {\sc ii} regions possibly dominate
the scale of structure in spiral arms.


\begin{chapthebibliography}{}
\bibitem{GDM01} Gaensler, B. M., Dickey, J. M., McClure-Griffiths, N. M.,
           Green, A. J., Wieringa, M. H., \& Haynes, R. F. 2001, ApJ,
           549, 959
\bibitem{MGD01} McClure-Griffiths, N. M., Green, A. J., Dickey, J. M.,
           Gaensler, B. M., Haynes, R. F., \& Wieringa, M. H. 2001,
           ApJ, 551, 394 
\end{chapthebibliography}
\end{document}